\documentclass[showpacs,preprintnumbers,amsmath,amssymb,aps, prd]{revtex4}
\usepackage{graphicx}
\usepackage{grffile}
\usepackage{mathrsfs, mathtools}
\usepackage{caption}
\usepackage{float}
\usepackage{xcolor}
\usepackage{dcolumn}
\usepackage{bm}
\usepackage{graphicx,subfigure,epsfig}
\usepackage{multirow}
\newcommand{\lagr}{\mathscr{L}}
\begin{document}
\title{Shadow, quasinormal modes, greybody bounds, and Hawking sparsity of Loop Quantum Gravity motivated non-rotating black hole}
\author{Sohan Kumar Jha}
\email{sohan00slg@gmail.com}
\affiliation{Chandernagore College, Chandernagore, Hooghly, West
Bengal, India}

\date{\today}
\begin{abstract}
\begin{center}
Abstract
\end{center}
We consider Loop Quantum Gravity(LQG) motivated $4D$ polymerized black hole and study shadow, quasinormal modes, and Hawking radiation. We obtain analytical expressions of photonsphere radius and shadow radius and study their qualitative and quantitative nature of variation with respect to the LQG parameter $\alpha$. We also show shadows of the black hole for various values of $\alpha$. Our study reveals that both radii increase with an increase in the parameter value. We, then, study quasinormal modes for scalar and electromagnetic perturbations using the $6th$ order WKB method. Our study reveals that the LQG parameter impacts quasinormal modes. We observe that the oscillation of gravitational wave(GW) and decay rate decrease as $\alpha$ increases. At the same time, the error associated with the $6th$ order WKB method increases with an increase in $\alpha$. The ringdown waveform for electromagnetic and scalar perturbations is shown. We also study greybody bounds, power spectrum, and sparsity of Hawking radiation. Greybody bounds for electromagnetic perturbations do not depend on $\alpha$. For scalar perturbation, greybody bounds increase as the LQG parameter increases, but the variation with $\alpha$ is very small. The peak of the power spectrum as well as total power emitted decrease as we increase the value of $\alpha$. Also, the sparsity of Hawking radiation gets significantly impacted by quantum correction. Finally, we obtain the area spectrum of the black hole. It is found to be significantly different than that for the Schwarzschild black hole.
\end{abstract}
\maketitle
\section{Introduction}
The observation of shadows of supermassive black holes(BHs) $M87^{*}$ and $SgrA^{*}$ by event horizon telescope(EHT) \cite{ak2019, ak2022} has validated the remarkable accuracy of General Relativity(GR) given by Einstein \cite{ae1936}. The existence of BHs was predicted by GR. It was shown by Hawking and Penrose that BHs formed by the gravitational collapse of massive objects would eventually have a spacetime singularity \cite{sw1970}. The existence of such a singularity leads to the breakdown of physical laws and the divergence of scalar invariants. As a result, geodesics are incomplete. It is generally believed that no such singularity exists in nature and such singularities are unavoidable features of classical GR. Wheeler suggested that quantum gravity can help us resolve the spacetime singularity \cite{wj1963}. The first regular black hole(RBH) without singularity was proposed by Bardeen \cite{bj1968}. In Bardeen's black hole, we have de sitter-like region, resulting in a regular center. A significant number of studies have been devoted to studying various models of RBHs. Loop Quantum Gravity(LQG) is considered to be one of the viable models of the quantum theory of gravity \cite{aa2006, aa2007, vk2007}. LQG uses a non-perturbative technique and employs area and volume quantization to resolve singularity [\citenum{aa12006}-\citenum{cr1998}]. Research has till now been confined to spherically symmetric BHs due to the complexity involved in solving the entire system [\citenum{aa12006}-\citenum{gr2008}]. Discreteness of spacetime, suggested by LQG, is preserved by an elegant technique called phase space quantization or polymerization [\citenum{bc2007}-\citenum{gr2008}]. Based on works that have been done in this field [\citenum{aa12006}-\citenum{bc2008}], Peltola and Kunstatter \cite{ap2009} reported a static, spherically symmetric, single horizon, regular black hole. Unlike other RBHs, it has one horizon whereby the problem of mass inflation at the inner horizon is removed. At the same time, spacetime is hyperbolic globally and the geodesics are complete. Several studies have been made in non-rotating RBHs [\citenum{id1992}-\citenum{ak2020}] and rotating RBHs [\citenum{ak2020}-\citenum{ae2021}].\\
The shadow of a black hole is a manifestation of its strong gravitation field. Black hole shadow has been a topic of intense research for quite some time now. The first image of the shadow of a supermassive black hole $M87^{*}$ has only increased the interest of the research community in studying various aspects of shadow. Bardeen, Press, and Teukolsky studied the shadow of a Kerr BH in \cite{JM}. The shadow of the Schwarzschild black was studied by Synge in \cite{JL}. The shadow of a BH surrounded by a bright accretion disk was studied by Luminet \cite{JP}. Narayan in his article \cite{RN} studied the shadow of a spherically accreting black hole. In article \cite{YG}, authors have studied the shadow and photon rings of Reissner-Nordstrom(RN). A several studies have been made in [\citenum{RC}-\citenum{SN}] to use shadow for detection of dark matter.\\
Quasinormal modes are one of the significant aspects of black hole physics. They are related to the emission of GWs from perturbed BHs that eventually die down due to dissipation [\citenum{CV}-\citenum{CHANDRA}]. These modes are called quasinormal as they are transient. Quasinormal modes are complex numbers where the real part corresponds to the frequency of the GW and the imaginary part gives the decay rate. There are three phases that BHs experience after perturbation: inspiral, merger, and ringdown. Quasinormal modes, for remnant BHs, are related to the ringdown phase. Quasinormal modes bear the signature of the underlying spacetime. Thus, it is important to study quasinormal modes to gauge the impact of quantum correction. Several articles have been devoted to studying quasinormal modes of various BHs [\citenum{CM}-\citenum{YY1}]. Another important phenomenon related to a BH is Hawking radiation. It was Hawking who showed that BHs emit radiation \cite{HAWKING}. Hawking took into account quantum consequences to prove it. When a pair production occurs close to the event horizon, one of the particles enters the BH and the second particle moves away from BH. It is the second particle that forms the Hawking radiation [\citenum{HH}-\citenum{HC}]. A number of different ways can be employed to obtain Hawking temperature [\citenum{SW}-\citenum{SI}]. The greybody factor gives the probability of Hawking radiation reaching an asymptotic observer. Thus, it is an important quantity. Matching method [\citenum{SF}-\citenum{JE}] or WKB method \cite{MK, CH} can be used to calculate greybody bounds. Visser \cite{GB} gave an elegant method to find greybody bounds. This method has been used in \cite{GB1, WJ}. \\
This manuscript is organized as follows. In Sec. (II), we introduce the LQG motivated $4D$ polymerized black hole metric and obtain analytical expressions of radii of photonsphere and shadow. We also study the qualitative and quantitative variation of those radii with respect to the LQG parameter $\alpha$. In the next section, we study the quasinormal modes of the black hole for scalar and electromagnetic perturbations and probe the effect of the LQG parameter on quasinormal modes. In Sec. (IV), we obtain the analytical expressions of Hawking temperature and greybody bounds and investigate the effect of quantum correction on them. In Sec. (V), we study the power spectrum and sparsity of Hawking radiation. In Sec. (VI), we obtain the area spectrum of the LQG-motivated black hole. We conclude our article in Sec. (VII) with a brief discussion of our results. Throughout the paper, we use $G=c=M=\hbar=1$.
\section{LQG motivated non-rotating black hole and its shadow}
\begin{figure}[H]
\centering
\includegraphics[width=0.4\columnwidth]{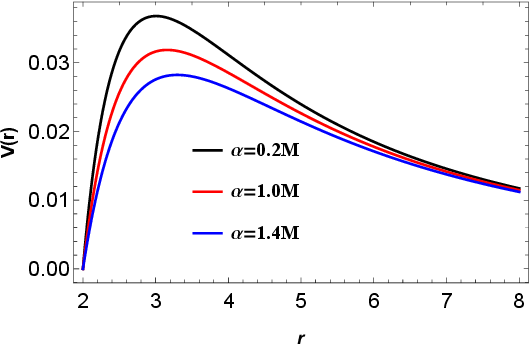}
\caption{Potential for various values of $\alpha$. Here, we have taken $\mathcal{L}$=1.}
\label{potential}
\end{figure}

\begin{figure}[H]
\centering
\subfigure[]{
\label{vfig1}
\includegraphics[width=0.4\columnwidth]{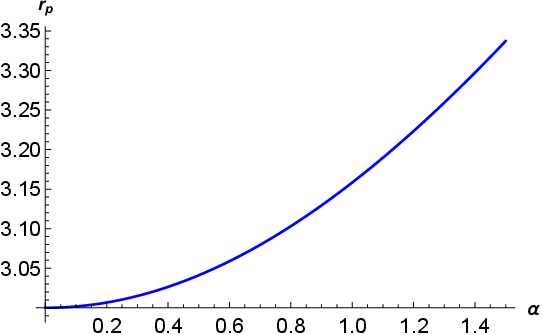}
}
\subfigure[]{
\label{vfig2}
\includegraphics[width=0.4\columnwidth]{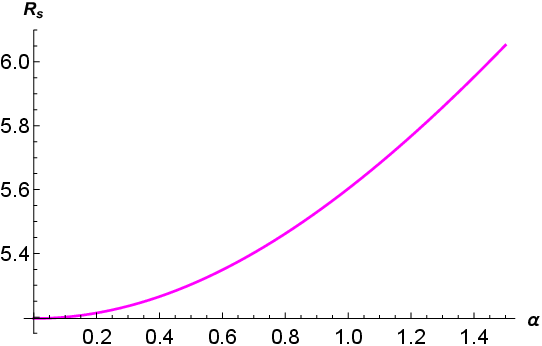}
}
\caption{Variation of $r_p$ and $R_s$ against $\alpha$.}
\label{radius}
\end{figure}

\begin{figure}[H]
\centering
\includegraphics[width=0.4\columnwidth]{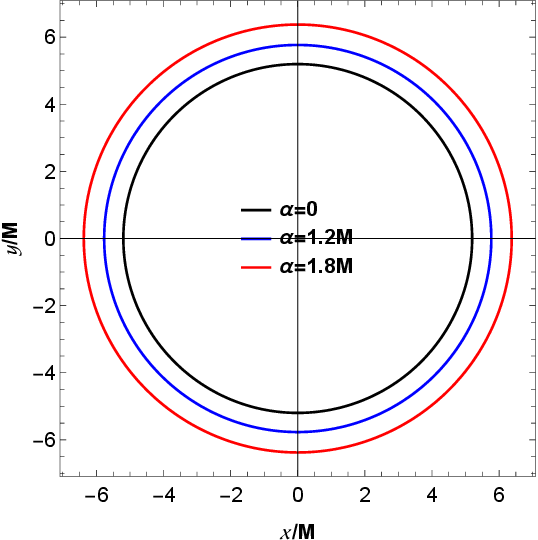}
\caption{Shadow for various values of $\alpha$.}
\label{shadow}
\end{figure}

Peltola and Kunstatter, with the help of effective field theory technique, derived the following LQG motivated $4D$ polymerized static and spherically symmetric black hole metric \cite{ap2009}
\begin{eqnarray}\label{metric}
ds^{2}=-\left(\sqrt{1-\frac{\alpha^2}{z^2}}-\frac{2M}{z}\right)dt^{2}+\frac{\left(1-\frac{\alpha^2}{z^2}\right)^{-1}}{\left(\sqrt{1-\frac{\alpha^2}{z^2}}-\frac{2M}{z}\right)}dz^{2}+ z^2 (d\theta^{2}+\sin^{2}\theta d\phi^{2}).
\end{eqnarray}
The above metric is singular at $\bar{y}=\alpha$. This singularity can be removed by using the transformation $z=\sqrt{r^2+\alpha^2}$. With this transformation, the metric (\ref{metric}) becomes
\begin{eqnarray}\label{metric2}
ds^{2}= -\left(\frac{r-2M}{\sqrt{r^2+\alpha^2}}\right)dt^{2}+\frac{1}{\left(\frac{r-2M}{\sqrt{r^2+\alpha^2}}\right)}dr^{2}
+(r^2+\alpha^2)(d\theta^{2}+\sin^{2}\theta d\phi^{2}).
\end{eqnarray}
The above metric reduces to the Schwarzschild metric when we put $\alpha=0$. Here, the range of the radial coordinate $r$ is $0\leq r\leq \infty$. The event horizon of the black hole represented by the metric (\ref{metric2}) is located at $r=2M$, irrespective of the value of $\alpha$. Ricci scalar R and Kretschmann scalar K for the metric (\ref{metric2}) are given by
\begin{eqnarray}\nonumber
\hspace*{-7cm}
R&=&\frac{2 \alpha ^2 \left(3 M+\sqrt{\alpha ^2+r^2}\right)-2 r^3+2 r^2 \sqrt{\alpha ^2+r^2}-5 \alpha ^2 r}{\left(\alpha ^2+r^2\right)^{5/2}},\\\nonumber
K&=&\frac{\splitfrac{4 \alpha ^6+12 M^2 \left(3 \alpha ^4+4 r^4-4 \alpha ^2 r^2\right)+4 M r \left(-15 \alpha ^4-4 r^4+14 \alpha ^2 r^2\right.}{\left.+4 \alpha ^2 r \sqrt{\alpha ^2+r^2}+4 r^3 \sqrt{\alpha
^2+r^2}\right)+8 r^6+12 \alpha ^2 r^4+41 \alpha ^4 r^2-8 r^5 \sqrt{\alpha ^2+r^2}-8 \alpha ^2 r^3 \sqrt{\alpha ^2+r^2}}}{\left(\alpha ^2+r^2\right)^5}.
\end{eqnarray}
These expressions reveal that the scalar invariants are finite everywhere. It implies that the metric (\ref{metric2}) represents a regular spacetime globally and geodesics are complete.\\
We next study null geodesics in the background of LQG motivated black hole given by the metric (\ref{metric2}). As the black hole we are considering is spherically symmetric, we, without loss of generality, can consider the equatorial plane with $\theta=\frac{\pi}{2}$. For equatorial plane, the ansatz (\ref{metric2}) reduces to
\begin{equation}\label{metric3}
ds^2=-f(r)dt^2+\frac{dr^2}{f(r)}+ h(r)d\phi^2,
\end{equation}
where $f(r)=\frac{r-2M}{\sqrt{r^2+\alpha^2}}$ is the lapse function and $h(r)=r^2+\alpha^2$. As the polymerized black hole is static and spherically symmetric, the energy $\mathcal{E}=-p_{\mu} \xi_{(t)}^{\mu}$ and the angular momentum $\mathcal{L}=p_{\mu} \xi_{(\phi)}^{\mu}$ along the geodesics are conserved. Here, $\xi_{(t)}^{\mu}$ and ${\xi_{(\phi)}^{\mu}}$ are the Killing vectors due to time-translational and rotational invariance respectively \cite{sc1975}. Thus, $\mathcal{E}=-p_t$ is the energy of a photon and $\mathcal{L}=p_\phi$ is the angular momentum. The expressions of $p_t$ and $p_\phi$ are obtained from the Lagrangian corresponding to the metric (\ref{metric3}). The Lagrangian is given by
\begin{equation}
\lagr=-f(r)\dot{t}^2+\frac{\dot{r}^2}{f(r)}+h(r)\dot{\phi}^2.
\end{equation}
Now, we have $p_q=\frac{\partial \lagr}{\partial \dot{q}}$, where $p_q$ is the conjugate momentum to the coordinate $q$. With the help of this definition, we obtain
\begin{eqnarray}\nonumber
p_t&=&\frac{\partial \lagr}{\partial \dot{t}}=-f(r)\dot{t}, \\\nonumber
p_r&=&\frac{\partial \lagr}{\partial \dot{r}}=\frac{\dot{r}}{f(r)}, \\
p_\phi&=&\frac{\partial \lagr}{\partial \dot{\phi}}=h(r)\dot{\phi}.
\end{eqnarray}
Here, the dot is differentiation with respect to an affine parameter $\tau$. Differential equations of motion are
\begin{equation}
\frac{d t}{d \tau}=\frac{\mathcal{E}}{f(r)} \quad \text{and} \quad \frac{d \phi}{d \tau}=\frac{\mathcal{L}}{h(r)}.\label{Conserved}
\end{equation}
Using Eq.(\ref{Conserved}) and Eq.(\ref{metric3}), we obtain the differential equation for the null geodesics as
\begin{equation}
\left(\frac{d r}{d \tau}\right)^{2} \equiv \dot{r}^{2}
=\mathcal{E}^{2}-V(r),
\end{equation}
where $V(r)$ is the potential given by
\begin{equation}
V(r)=\frac{\mathcal{L}^{2} f(r)}{h(r)}.
\end{equation}
The unstable circular photon orbits are located at the peak of the above potential. In Fig. (\ref{potential}), we plot the potential with respect to r for various values of $\alpha$. We observe that the peak of the potential shifts towards the right as we increase the value of the parameter $\alpha$. It implies that the photon radius increases as we increase the value of the parameter. This finding will be confirmed in subsequent results.
For circular photon orbits of radius $r_{p}$, we must have
\begin{equation}
V(r_p)=0,\quad\frac{d V}{d r}|_{r=r_{p}}=0,\quad\text{and}\quad \frac{\partial^2V}{\partial r^2}|_{r=r_{p}}<0.
\label{condition}
\end{equation}
The middle equation yields
\begin{equation}
\frac{f^{\prime}(r_p)}{f(r_p)}=\frac{h^{\prime}(r_p)}{h(r_p)}.
\end{equation}
The above equation, on simplification, produces
\begin{equation}
2r^2-6Mr-\alpha^2=0.
\end{equation}
We obtain two roots from the above equation. Using the first and third conditions from Eq. (\ref{condition}), we exclude one solution and retain the other. The radius of the photonsphere is given by $r_p=\frac{1}{2} \left(\sqrt{2 \alpha^2+9 M^2}+3 M\right)$. This analytical expression shows that the LQG parameter $\alpha$ has a significant impact on the radius of the photonsphere. For distant observers, the shadow radius is equal to the impact parameter. Thus, the shadow radius is given by
\begin{eqnarray}\nonumber
b_p=R_{s}&=&\frac{\mathcal{L}}{\mathcal{E}}=\sqrt{\frac{h(r_p)}{f(r_p)}}\\
&=&\frac{3^{3/4} \sqrt{\frac{\left(\alpha ^2+M \left(\sqrt{2 \alpha ^2+9 M^2}+3 M\right)\right)^{3/2}}{\sqrt{2 \alpha ^2+9 M^2}-M}}}{\sqrt[4]{2}}.
\end{eqnarray}
In the limit $\alpha\rightarrow 0$, we get the values of photon radius and shadow radius for the Schwarzschild black hole i.e., $r_p=3M$ and $R_s=3\sqrt{3}M$. To understand the qualitative nature of variation of photon and shadow radius with respect to LQG parameter, we plot $r_p$ and $R_s$ against $\alpha$ in Fig. (\ref{radius}). Quantitative values of photon radius $r_p$ and shadow radius $R_{s}$ are given in Table (\ref{shadowradius}) for different values of the black hole parameter.
\begin{table}[!htp]
\centering
\caption{Various values of photon radius and shadow radius for different values of $\alpha$.}
\begin{tabular}{|c|c|c|c|c|c|c|c|c|c|c|c|}
\hline
$\alpha$ & 0 & 0.2 & 0.4 & 0.6 & 0.8 & 1.0 & 1.2 & 1.4 & 1.6 & 1.8 & 2.0 \\
\hline
$r_p$ & 3.0 & 3.00665 & 3.02643 & 3.05885 & 3.10312 & 3.15831 & 3.22337 & 3.29722 & 3.37883 & 3.46723 & 3.56155 \\
\hline
$ R_s$ & 5.19615 & 5.21343 & 5.26468 & 5.34832 & 5.46193 & 5.60259 & 5.76717 & 5.95258 & 6.15592 & 6.3746 & 6.60632 \\
\hline
\end{tabular}
\label{shadowradius}
\end{table}
One observation we can make from Fig. (\ref{radius}) and Table (\ref{shadowradius}) is that both the radii increase as we increase the value of the parameter $\alpha$. The variation of photon radius and shadow radius is significant with respect to the parameter $\alpha$. To study the shadow of the black hole given by (\ref{metric2}), we use two celestial coordinates:
$$
\begin{aligned}
& x=\lim _{r_o \rightarrow \infty}\left[-\left.r_o^2 \sin \theta_o \frac{d \phi}{d r}\right|_{\theta-\theta_o}\right], \\
& y=\lim _{r_o \rightarrow \infty}\left[\left.r_o^2 \frac{d \theta}{d r}\right|_{\theta-\theta_o}\right],
\end{aligned}
$$
where $\left(r_o, \theta_o\right)$ is the observer's position at infinity. For an observer in the equatorial plane i.e., $\theta_0=\pi / 2$, we have
$$
R_s \equiv \sqrt{x^2+y^2}.
$$
Shadows for the black hole are shown below for various values of $\alpha$. For $\alpha=0$, we obtain the shadow for the Schwarzschild black hole. For non-zero values of the parameter, we obtain the shadow for the quantum-corrected black hole. Fig. (\ref{shadow}) shows that the quantum correction has an observable impact on the shadow. From the plot, we also observe that the shadow radius increases with the parameter $\alpha$.
This concludes our discussion of photonsphere and shadow for the LQG-motivated $4D$ polymerized black hole.
\section{Quasinormal modes of non-rotating Simpson-Visser black hole}
In this section, We investigate quasinormal modes of LQG motivated $4D$ polymerized black hole for scalar and electromagnetic perturbations. Here, the impact of the scalar field or the electromagnetic field on the background spacetime is considered to be negligible. To study quasinormal modes, we first consider the equation for the relevant field and then, reduce it to a Schr$\ddot{o}$dinger-like equation. For the scalar field, we will have the Klein-Gordon equation and for the electromagnetic field, we will consider Maxwell equations. For the massless scalar field, we have
\begin{eqnarray}
\frac{1}{\sqrt{-g}}{\partial \mu}(\sqrt{-g}g^{\mu\nu} \partial_{\nu}\psi) =0,
\label{scalar}
\end{eqnarray}
and for the electromagnetic field, we have
\begin{equation}
\frac{1}{\sqrt{-g}}{\partial \nu }(F_{\rho\sigma}g^{\rho\mu}g^{\sigma\nu}\sqrt{-g})=0,
\label{em}
\end{equation}
where $ F_{\rho\sigma}={\partial \rho}A^\sigma-{\partial \sigma}A^\rho $, $A_\nu$ being electromagnetic four-potential.
We now introduce the tortoise coordinate:
\begin{eqnarray}
\text{d}r_*=\frac{\text{d}r}{f(r)}.
\label{tortoise}
\end{eqnarray}
With the help of tortoise coordinate, Eqs.(\ref{scalar}) and (\ref{em}) reduce to the following Schr$\ddot{o}$dinger-like form
\begin{equation}
-\frac{\text{d}^2\phi}{\text{d}{r^2_*}}+V_{\text{eff}}(r) \phi=\omega ^{2}\phi,
\label{schrodinger}
\end{equation}
where the effective potential is given by
\begin{eqnarray}
V_{\text{eff}}(r)&=&\frac{(1-s^2)f(r)}{r}\frac{\text{d}f(r)}{\text{d}r}+\frac{f(r)\ell(\ell+1)}{r^{2}}\\\nonumber&=&\frac{(r-2 M) \left(r^2 \left(\ell (\ell+1) \sqrt{\alpha ^2+r^2}-2 M \left(s^2-1\right)\right)+\alpha ^2 \left(\ell (\ell+1) \sqrt{\alpha ^2+r^2}-r s^2+r\right)\right)}{r^2 \left(\alpha^2+r^2\right)^2}.
\label{vtotal}
\end{eqnarray}
Here, $\ell$ is the angular momentum and s is the spin. For $s=0$, we obtain the effective potential for scalar perturbation, and for $s=1$, we obtain the effective potential for electromagnetic perturbation. Since the effective potential influences quasinormal modes, we briefly study the variation of the effective potential for various scenarios.
\begin{figure}[H]
\centering
\subfigure[]{
\label{vfig1}
\includegraphics[width=0.4\columnwidth]{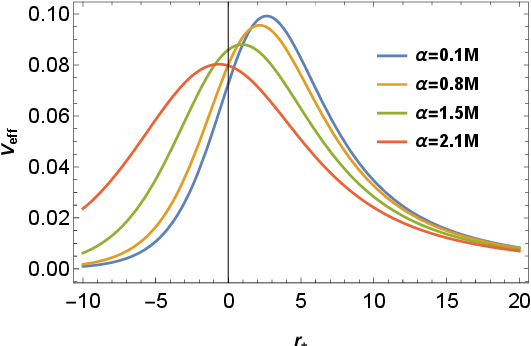}
}
\subfigure[]{
\label{vfig2}
\includegraphics[width=0.4\columnwidth]{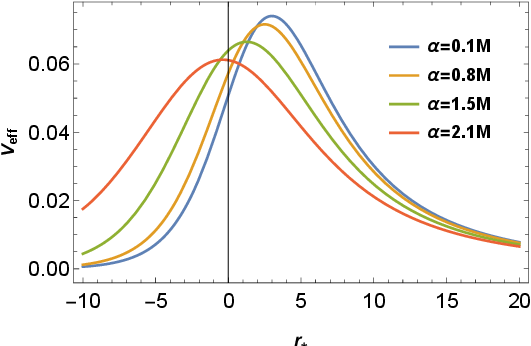}
}
\subfigure[]{
\label{vfig3}
\includegraphics[width=0.4\columnwidth]{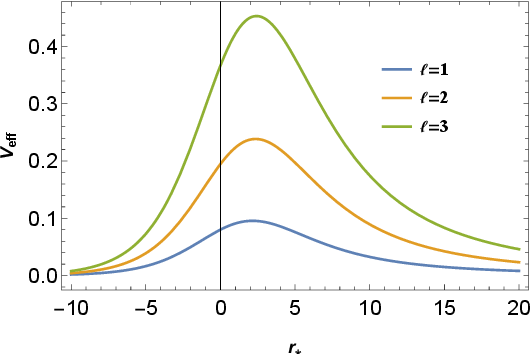}
}
\subfigure[]{
\label{vfig4}
\includegraphics[width=0.4\columnwidth]{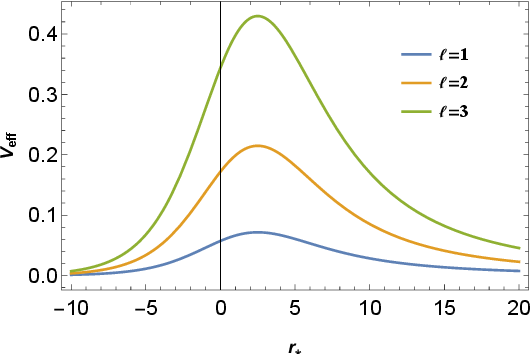}
}
\caption{Variation of effective potential with respect to tortoise coordinate $r_*$. The upper ones are for various values of $\alpha$ with $\ell=1$ and the lower ones are for various values of angular momentum with $\alpha=0.8M$. The left ones are for scalar perturbations and the right ones are for electromagnetic perturbations.}
\label{vfig}
\end{figure}
From Fig. (\ref{vfig}) we observe that the peak of the potential increases as we increase the angular momentum but decreases with the increase in the parameter $\alpha$. We also see that the position of the peak shifts towards the right as we increase the angular momentum $\ell$ or decrease the parameter $\alpha$.\\
Next, with the help of the $6th$ order WKB method, we obtain quasinormal modes. WKB method to calculate quasinormal modes was first developed by Schutz and Will \cite{schutz}. It was later extended to higher orders \cite{iyer, iyer1, konoplya1}. The $6th$ order WKB method yields the following expression of quasinormal frequencies:
\begin{equation}
\frac{\text{i}(\omega ^{2}-V_{0})}{\sqrt{-2V_{0}^{''}}}-\sum ^{6}_{\text{i}=2}\Omega_\text{i}=n+\frac{1}{2},
\label{WKB}
\end{equation}
where $V_{0}$ and $V''_{0}$ represent the height of the effective potential and the second derivative with respect to the tortoise coordinate of the potential at its maximum, respectively. $\Omega_\text{i}$ are the correction terms given in \cite{schutz, iyer, iyer1, konoplya1}. With the help of Eq.(\ref{WKB}), we calculate some of the values of quasinormal frequencies of scalar and electromagnetic perturbations for various values of angular momentum $\ell$ and parameter $\alpha$. In Table \ref{QNMS}, we show numerical values of quasinormal modes of scalar perturbation for different values of angular momentum $\ell$ and parameter $\alpha$ keeping overtone number $n=0$. In Table \ref{QNMEM}, we show quasinormal modes of electromagnetic perturbation for different values of angular momentum and LQG parameter keeping overtone number $n=0$. We also calculate the error associated with the $6th$ order WKB method defined by
\begin{equation}
\Delta_6=\frac{|\omega_7 -\omega_5|}{2},
\end{equation}
where $\omega_5$ and $\omega_7$ are quasinormal frequencies obtained using $5th$ order and $7th$ order WKB method respectively.
\begin{table}[!htp]
\centering
\caption{Quasinormal frequencies for scalar field with $n=0$.}
\setlength{\tabcolsep}{1mm}
\begin{tabular}{|c|c|c|c|c|c|c|}
\hline
$\alpha $ & $\ell=1$ & $\Delta _6$ & $\ell$=2 & $\Delta _6$ & $\ell$=3 & $\Delta _6$ \\
\hline
0 & 0.367984\, -0.0944086 i & 0.0000230659 & 0.532369\, -0.0953138 i & $3.28709\times 10^{-6}$ & 0.711038\, -0.0957039 i &
$7.63696\times 10^{-7}$ \\
\hline
0.25 & 0.367427\, -0.0940281 i & 0.0000260222 & 0.531502\, -0.0949261 i & $4.15212\times 10^{-6}$ & 0.709848\, -0.0953155 i &
$9.94250\times 10^{-7}$ \\
\hline
0.5 & 0.365787\, -0.0929174 i & 0.0000427638 & 0.528953\, -0.0937928 i & $6.94819\times 10^{-6}$ & 0.706351\, -0.0941796 i &
$1.64054\times 10^{-6}$ \\
\hline
0.75 & 0.363152\, -0.0911605 i & 0.0000688479 & 0.524868\, -0.0919963 i & 0.0000112364 & 0.700755\, -0.0923778 i & $2.63228\times 10^{-6}$ \\
\hline
1.0 & 0.359653\, -0.0888735 i & 0.0000987704 & 0.519462\, -0.0896536 i & 0.0000155716 & 0.693357\, -0.0900269 i & $3.58075\times 10^{-6}$ \\
\hline
1.25 & 0.355447\, -0.0861843 i & 0.000124503 & 0.512981\, -0.0868957 i & 0.0000178107 & 0.684502\, -0.087259 i & $3.94790\times 10^{-6}$ \\
\hline
\end{tabular}
\label{QNMS}
\end{table}
\begin{table}[!htp]
\centering
\caption{Quasinormal frequencies for electromagnetic field with $n=0$.}
\setlength{\tabcolsep}{1mm}
\begin{tabular}{|c|c|c|c|c|c|c|}
\hline
$\alpha $ & $\ell=1$ & $\Delta _6$ & $\ell$=2 & $\Delta _6$ & $\ell$=3 & $\Delta _6$ \\
\hline
0 & 0.248191\, -0.092637 i & 0.000142597 & 0.457593\, -0.0950112 i & $7.04358\times 10^{-6}$ & 0.656898\, -0.0956171 i &
$1.13864\times 10^{-6}$ \\
\hline
0.25 & 0.247877\, -0.0922998 i & 0.000170868 & 0.456865\, -0.094625 i & $8.85698\times 10^{-6}$ & 0.655805\, -0.0952289 i &
$1.45442\times 10^{-6}$ \\
\hline
0.5 & 0.246938\, -0.0913156 i & 0.000284307 & 0.454725\, -0.0934984 i & 0.0000147938 & 0.652596\, -0.0940947 i & $2.4338\times 10^{-6}$ \\
\hline
0.75 & 0.245407\, -0.0897485 i & 0.000462204 & 0.4513\, -0.0917176 i & 0.0000236744 & 0.647466\, -0.0922994 i & $3.88534\times 10^{-6}$ \\
\hline
1.0 & 0.24337\, -0.0876629 i & 0.000642422 & 0.446776\, -0.0894034 i & 0.0000323796 & 0.640697\, -0.0899635 i & $5.2812\times 10^{-6}$ \\
\hline
1.25 & 0.240962\, -0.0851176 i & 0.000710422 & 0.441366\, -0.0866884 i & 0.0000357733 & 0.632612\, -0.087222 i & $5.83228\times 10^{-6}$ \\
\hline
\end{tabular}
\label{QNMEM}
\end{table}
From Tables (\ref{QNMS},\ref{QNMEM}), we can infer that the real part of quasinormal frequencies decreases with an increase in parameter value $\alpha$ for a particular value of $\ell$. Additionally, it is observed for both perturbations that the real part of quasinormal modes increases as we increase the angular momentum $\ell$. We can observe from the Tables (\ref{QNMS},\ref{QNMEM}) that the decay rate or damping rate increases as we decrease the value of parameter $\alpha$ or increase the angular momentum for both perturbations. It is also observed that the error associated with the $6th$ order WKB method increases with an increase in the LQG parameter value $\alpha$. Next, we investigate the qualitative nature of variation of quasinormal frequency for various aspects.
\begin{figure}[H]
\centering
\subfigure[]{
\label{qnmimfig1}
\includegraphics[width=0.4\columnwidth]{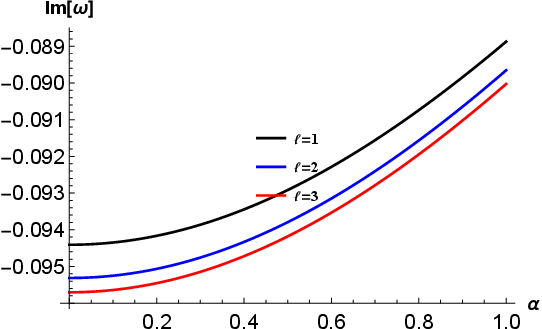}
}
\subfigure[]{
\label{qnmimfig2}
\includegraphics[width=0.4\columnwidth]{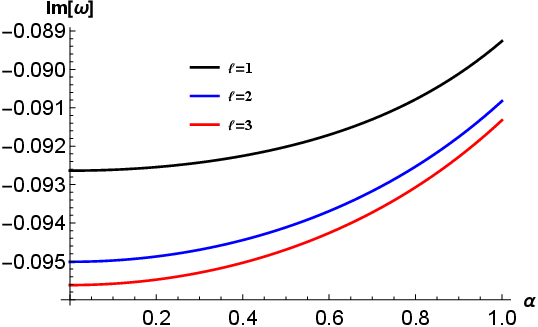}
}
\caption{It gives the variation of the imaginary part of the quasinormal frequency with respect to $\alpha$ for various values of $\ell$. The left one is for the scalar field and the right one is for the electromagnetic field.}
\label{qnmimfig}
\end{figure}
\begin{figure}[H]
\centering
\subfigure[]{
\label{qnmrefig1}
\includegraphics[width=0.4\columnwidth]{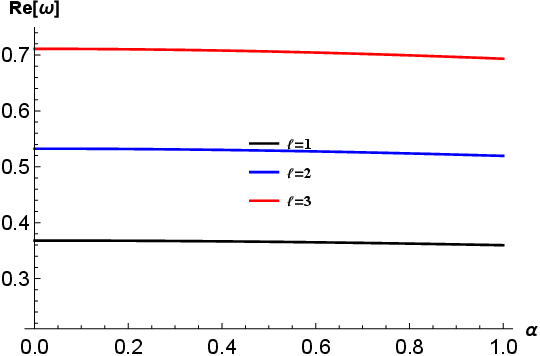}
}
\subfigure[]{
\label{qnmrefig2}
\includegraphics[width=0.4\columnwidth]{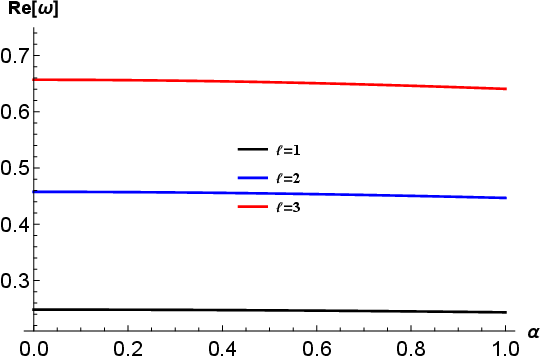}
}
\caption{It gives the variation of the real part of the quasinormal frequency with respect to $\alpha$ for various values of $\ell$. The left one is for the scalar field and the right one is for the electromagnetic field.}
\label{qnmrefig}
\end{figure}
\begin{figure}[H]
\centering
\subfigure[]{
\label{qnmrefig1}
\includegraphics[width=0.4\columnwidth]{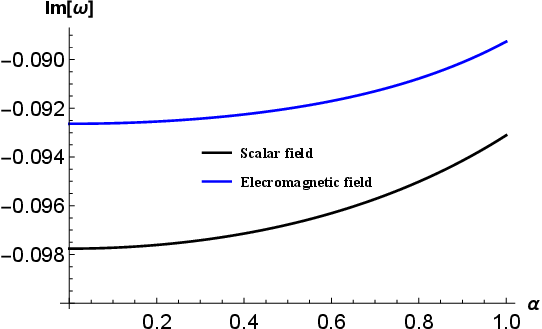}
}
\subfigure[]{
\label{qnmrefig2}
\includegraphics[width=0.4\columnwidth]{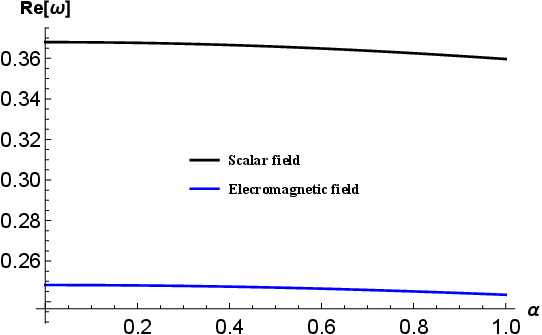}
}
\caption{Left one gives the variation of the imaginary part of the quasinormal frequency with respect to $\alpha$ for scalar and electromagnetic fields and the right one gives that for the real part. Here, we have taken $\ell=1$.}
\label{qnmimrefig}
\end{figure}
Fig.(\ref{qnmimfig}) and Fig.(\ref{qnmrefig}) reinforce the findings we have drawn from Tables (\ref{QNMS}, \ref{QNMEM}). We can also observe from the Fig. (\ref{qnmimrefig}) that the real part of quasinormal modes is larger for scalar perturbation, whereas, the imaginary part is larger for electromagnetic perturbation. It implies that the damping rate or decay rate is larger for scalar perturbation. We next study the convergence of the WKB method for various values of $(n,\ell)$ pair.
\begin{figure}[H]
\centering
\subfigure[]{
\label{qnmorderfig3}
\includegraphics[width=0.4\columnwidth]{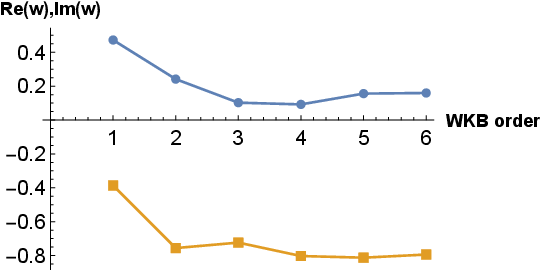}
}
\subfigure[]{
\label{qnmorderfig4}
\includegraphics[width=0.4\columnwidth]{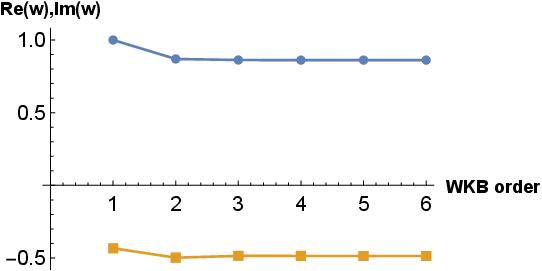}
}
\caption{Variation of the real and imaginary part of quasinormal frequencies with respect to WKB order for various values of $(n,\ell)$ pair. The left one is for $(3,0)$ pair and the right one is for $(2,4)$ pair. In each plot, the blue line is for the real part and the orange line is for the imaginary part of the quasinormal mode. Here, we have taken $\alpha=0.4M$.}
\label{qnmorderfig}
\end{figure}
From the above figure (\ref{qnmorderfig}) we observe that quasinormal frequencies fluctuate even for higher order when we consider the pair $(3,0)$. This confirms the finding in the article \cite{kd} where it is observed that WKB approximation is reliable when the angular momentum is high and the overtone number is low.
\section{Ringdown waveform}
\begin{figure}[H]
\centering
\subfigure[]{
\label{ringingscalar1}
\includegraphics[width=0.45\columnwidth]{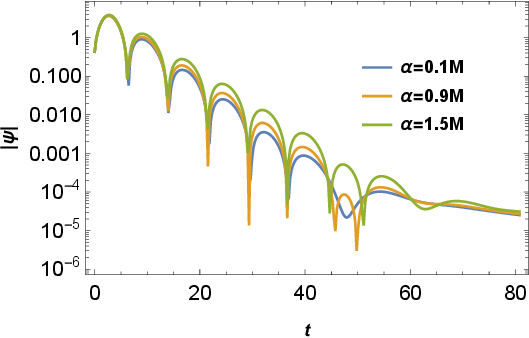}
}
\subfigure[]{
\label{ringingscalar2}
\includegraphics[width=0.45\columnwidth]{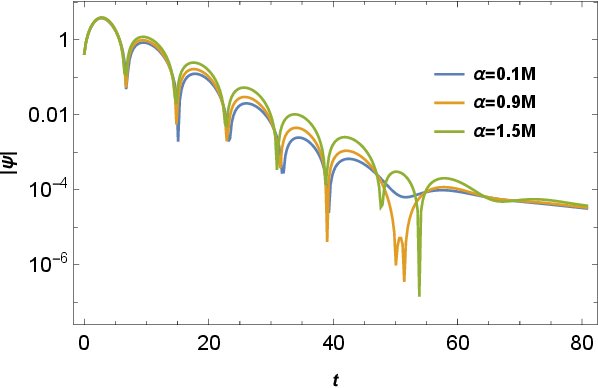}
}
\caption{Time domain profile for various values of $\alpha$. Left one is for scalar perturbation and the right one is for electromagnetic perturbation. Here, we have taken $\ell=2$.}
\label{ringingalpha}
\end{figure}

\begin{figure}[H]
\centering
\subfigure[]{
\label{ringingem1}
\includegraphics[width=0.45\columnwidth]{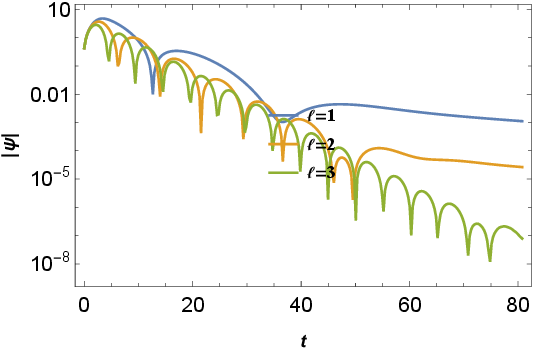}
}
\subfigure[]{
\label{ringingem2}
\includegraphics[width=0.45\columnwidth]{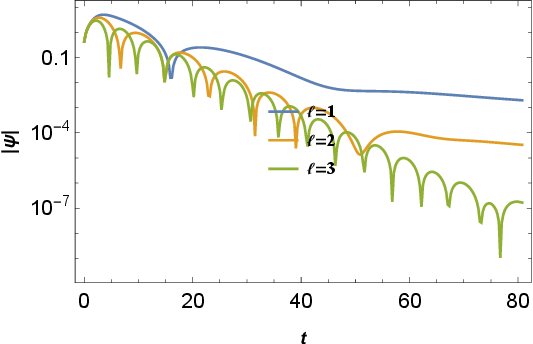}
}
\caption{Time domain profile for various values of $\ell$. Left one is for scalar perturbation and the right one is for electromagnetic perturbation. Here, we have taken $\alpha=0.8M$.}
\label{ringingl}
\end{figure}

In this section, we study the time evolution of the scalar and electromagnetic perturbations. For this purpose, we numerically solve the time-dependent wave equation using the time domain integration method given by Gundlach et al. in their article \cite{gundlach1} using initial conditions $\psi(r_*,t) = \exp \left[ -\dfrac{(r_*-\hat{r}_{*})^2}{2\sigma^2} \right]$ and $\psi(r_*,t)\vert_{t<0} = 0$, where we have taken $r_*=5$, $\hat{r}_*=0.4$. The values of $\Delta t$ and $\Delta r_{*}$ are taken such that the Von Neumann stability condition, $\frac{\Delta t}{\Delta r_*} < 1$, is satisfied.\\
In Fig. (\ref{ringingalpha}) we provide the ringdown waveform for various values of the parameter $\alpha$ keeping $\ell=2$ and in Fig. (\ref{ringingl}), we provide the waveform for various values of $\ell$ keeping $\alpha=0.8M$. From Fig. (\ref{ringingalpha}) we can clearly conclude that the frequency decreases as we increase the parameter $\alpha$. It can also be inferred from the figure that the decay rate, given by the magnitude of the slope of the maxima in the log graph, decreases as we increase $\alpha$. From Fig. (\ref{ringingl}) we can conclude that the frequency as well as the decay rate increases as we increase $\ell$. These are consistent with the conclusions drawn from Tables (\ref{QNMS}, \ref{QNMEM}).

\section{Hawking temperature and Bounds of the Greybody factor}
\begin{figure}[H]
\centering
\subfigure[]{
\label{gbsfig1}
\includegraphics[width=0.4\columnwidth]{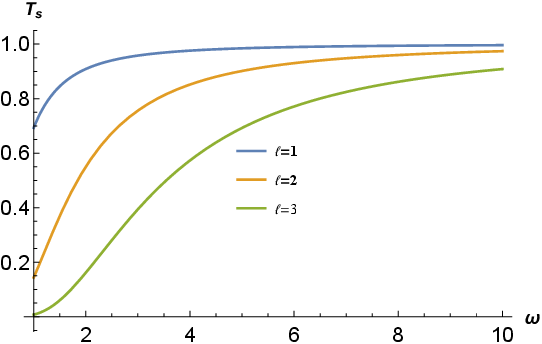}
}
\subfigure[]{
\label{gbemfig1}
\includegraphics[width=0.4\columnwidth]{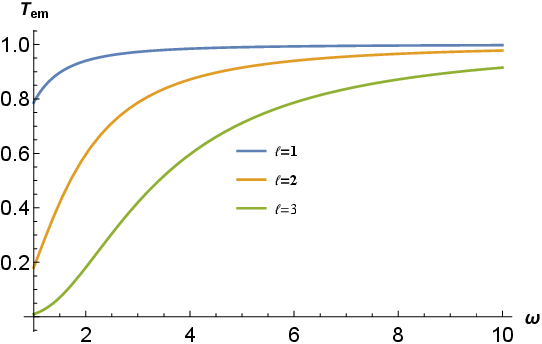}
}
\caption{Bounds of greybody factor for various values of $\ell$. Left one is for scalar perturbation and the right one is for electromagnetic perturbation. Here, we have taken $\alpha=0.6M$.}
\label{gbfig}
\end{figure}

\begin{figure}[H]
\centering
\includegraphics[width=0.5\columnwidth]{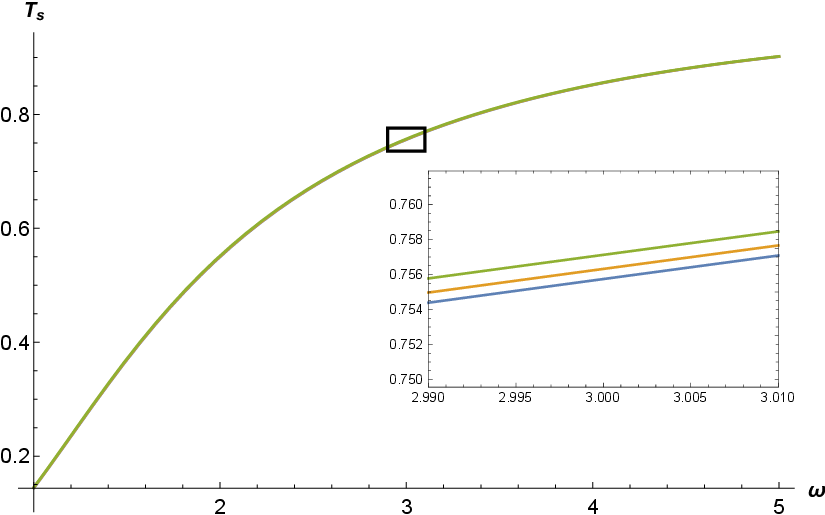}
\caption{Bounds of greybody factor for various values of $\alpha$. Within the frame, the green line is for $\alpha=1.6M$, the orange line is for $\alpha=1.0M$, and the blue line is for $\alpha=0.2M$. Here, we have taken $\ell=1$.}
\label{gbfig1}
\end{figure}

In this section, we intend to calculate the Hawking temperature and greybody bounds for the black hole under consideration. Hawking in his article \cite{HAWKING} showed that black holes emit radiation. That radiation is known as Hawking radiation. Bekenstein in his article \cite{BEK} and Keifer in his article \cite{KEIF} showed that it was necessary to associate a temperature with the horizon for consistency with thermodynamics. The Hawking temperature is given by
\begin{equation}
T_H=\frac{1}{4\pi \sqrt{-g_{tt}g_{rr}}}\frac{dg_{tt}}{dr}|_{r=r_h}.
\end{equation}
For the metric in consideration, we have $g_{tt}=-f(r)$ and $g_{rr}=\frac{1}{f(r)}$. Putting these values in the above equation, we get
\begin{equation}
T_H=\frac{1}{4 \pi \sqrt{\alpha ^2+4 M^2}}.
\label{hawking}
\end{equation}
The dependence of the Hawking temperature on the parameter $\alpha$ is evident from the above equation. We recover the value of the Hawking temperature for the Schwarzschild black hole if we put $\alpha=0$ in the above equation. To show the dependence graphically, we plot the Hawking temperature against $\alpha$.
\begin{figure}[H]
\centering
\subfigure[]{
\label{hw1}
\includegraphics[width=0.4\columnwidth]{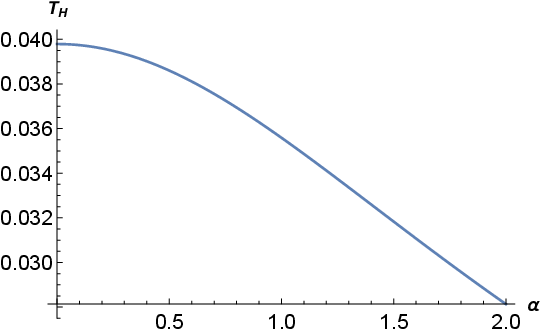}
}
\caption{Variation of Hawking temperature with respect to $\alpha$.}
\label{hwfig}
\end{figure}
We can observe that the Hawking temperature decreases as we increase the value of the parameter $\alpha$. The Hawking radiation observed by an asymptotic observer is different from the original radiation near the horizon of the black hole due to the redshift factor. Greybody distribution describes the Hawking radiation that is observed by an asymptotic observer. Here, we try to obtain the lower bound of the greybody factor for LQG motivated $4D$ polymerized black hole. A lot of research has been dedicated to bound greybody factor. Visser and Boonserm in their articles \cite{GB, GB1, GB2} gave an elegant way to lower bound the greybody factor. A rigorous bound of the transmission probability, which is the same as that of the graybody factor, is given by
\begin{equation}
T\geq sech^{2}(\frac{1}{2\omega}\int_{-\infty}^{\infty}|V_{\text{eff}}(r_*)|dr_*),\label{grey}
\end{equation}
where $r_*$ is the tortoise coordinate defined in Eq.(\ref{tortoise}) and $V_{\text{eff}}(r_*)$ is the potential given in Eq.(\ref{vtotal}). In terms of normal coordinate r, the above equation becomes
\begin{equation}
T\geq sech^{2}(\frac{1}{2\omega}\int_{r_h}^{\infty}|V_{\text{eff}}(r)|\frac{dr}{f(r)})\label{grey1}.
\end{equation}
If we use Eq.(\ref{vtotal}), then, the above equation reduces to
\begin{equation}
T\geq sech^2\left(\frac{-\frac{4 M \left(s^2-1\right) \left(1-\frac{2 M}{\sqrt{\alpha ^2+4 M^2}}\right)}{\alpha ^2}+\frac{2 \left(s^2-1\right) \left(\alpha \sqrt{\alpha ^2+4
M^2}-\left(\alpha ^2+4 M^2\right) \sinh ^{-1}\left(\frac{\alpha }{2 M}\right)\right)}{\alpha \left(\alpha ^2+4 M^2\right)}+\frac{\ell ^2}{M}+\frac{\ell }{M}}{4 \omega}\right).
\label{gb}
\end{equation}
The greybody bound for the scalar perturbation, $T_s$, is obtained when we put $s=0$ and the bound for the electromagnetic perturbation, $T_{em}$, is obtained by taking $s=1$. From Eq. (\ref{gb}) we see that $T_s$ depends on the LQG parameter $\alpha$ but $T_{em}$ is independent of it. The qualitative nature of variation of $T_s$ and $T_{em}$ are shown in Fig. (\ref{gbfig}) and Fig. (\ref{gbfig1}). In Fig. (\ref{gbfig}), we observe that the greybody bound decreases as we increase the angular momentum. It signifies that the probability of detecting Hawking radiation by an asymptotic observer decreases with $\ell$. Fig. (\ref{gbfig1}) shows that the greybody bound for scalar perturbation increases with the LQG parameter, but the amount of variation is small. It is also observed that the greybody bound approaches its maximum value of 1 faster for smaller values of angular momentum.
\section{Spectrum and sparsity of Hawking radiation}
In this section, we will study the effect of quantum correction on the spectrum and sparsity of Hawking radiation for both perturbations. Total power emitted as Hawking radiation by a black hole at Hawking temperature $T_{H}$ is given by \cite{yg2017, fg2016}
\begin{equation}
\frac{dE(\omega)}{dt}\equiv P_{tot} = \sum_\ell T (\omega)\frac{\omega}{e^{\omega/T_{H}}-1} \hat{k} \cdot \hat{n}~ \frac{d^3 k ~dA}{(2\pi)^3},
\end{equation}
where $dA$ is the surface element, $\hat{n}$ is unit normal to $dA$, and $T$ is the greybody factor given by Eq. (\ref{gb}). Since for massless particles we have $|k|=\omega$, the above equation for massless particles becomes
\begin{equation}\label{ptot}
P_{tot}=\sum_\ell \int_{0}^{\infty} P_\ell\left(\omega\right) d\omega.
\end{equation}
Here, $P_{\ell}$ is power spectrum in the $\ell th$ mode given by
\begin{equation}\label{pl}
P_\ell\left(\omega\right)=\frac{A}{8\pi^2}T(\omega)\frac{\omega^3}{e^{\omega/T_{H}}-1}.
\end{equation}
Although $A$ is a multiple of the horizon area, here, we have taken $A$ to be the horizon area as it will not affect the qualitative result \cite{yg2017}. Power spectrum $P_{\ell}(\omega)$ is important to study the sparsity of Hawking radiation. In Fig. (\ref{pl}), we study the qualitative nature of variation of power spectrum $P_{\ell}$ for different parameter values of the black hole. Here, we have plotted $P_{\ell}$ with respect to $\omega$ for various values of the LQG parameter $\alpha$.
\begin{figure}[H]
\centering
\subfigure[]{
\label{pls}
\includegraphics[width=0.4\columnwidth]{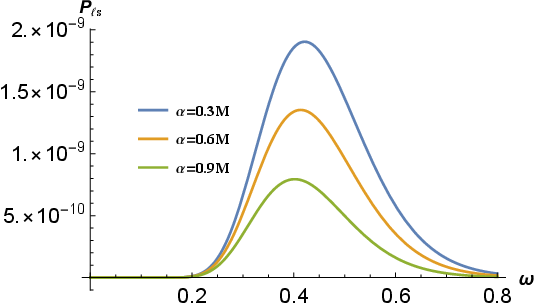}
}
\subfigure[]{
\label{plem}
\includegraphics[width=0.4\columnwidth]{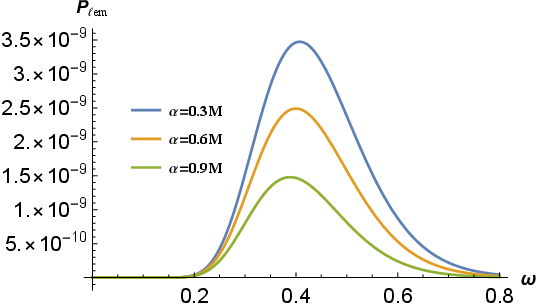}
}
\caption{Power spectrum of the black hole for various values of $\alpha$. Left one is for scalar perturbation and the right one is for electromagnetic perturbation. Here, we have taken $\ell=2$.}
\label{pl}
\end{figure}
We observe from Fig. (\ref{pl}) that for both perturbations, the maximum value of the power spectrum diminishes as we increase the value of $\alpha$ but the frequency, $\omega_{max}$, at which we have a maximum value of $P_{\ell}$ decreases with $\alpha$.\\
To have a better understanding of Hawking radiation emitted by black holes, we introduce a dimensionless parameter, $\eta$, that defines the sparsity of Hawking radiation as \cite{yg2017, fg2016, ac2020, sh2016, sh2015}
\begin{equation}
\eta=\frac{\tau_{gap}}{\tau_{emission}}.
\label{eta}
\end{equation}
Here, $\tau_{gap}$ is the average time gap between two successive radiation quanta. It is defined by
\begin{equation}\label{tgap}
\tau_{gap}=\frac{\omega_{max}}{P_{tot}}.
\end{equation}
The time that is taken by a radiation quantum for emission, $\tau_{emission}$, is defined by
\begin{equation}\label{temission}
\tau_{emission} \geq \tau_{localisation}=\frac{2 \pi}{\omega_{max}},
\end{equation}
where $\tau_{localisation}$ is the time period of the emitted wave of frequency $\omega_{max}$. Thus, we will have a continuous flow of Hawking radiation when $\eta\ll1$. A large value of $\eta$ signifies that the emission of radiation quanta is not continuous and the time gap between two radiation quanta is larger than the required time of radiation emission. The quantitative values of $\omega_{max}$, $P_{max}$, $P_{tot}$, and $\eta$, for scalar perturbations, are given in Table (\ref{sparsescalar}) and for electromagnetic perturbations, those values are given in Table (\ref{sparseem}). From these tables, we observe that the peak of the power spectrum and total power emitted decrease as we increase the value of the LQG parameter for both types of perturbations. On the other hand, the sparsity increases as we increase the value of $\alpha$. It means that the time interval between two radiation quanta increases as we increase the LQG parameter. If we compare the values of Table (\ref{sparsescalar}) and Table (\ref{sparseem}), we can infer that the sparsity of black hole is larger for electromagnetic perturbation. Since the variation of sparsity is significant for both perturbations, Hawking radiation may be used in the future to validate LQG.
\begin{table}[!htp]
\centering
\caption{Numerical values of $\omega_{max}$, $P_{max}$, $P_{tot}$, and $\eta$ for scalar perturbation for various values of $\alpha$ for $\ell=1$ mode.}
\setlength{\tabcolsep}{-.2mm}
\begin{tabular}{|c|c|c|c|c|c|c|c|c|c|}
\hline
$\alpha$ & 0.2 & 0.4 & 0.6 & 0.8 & 1.0 & 1.2 & 1.4 & 1.6 & 1.8 \\
\hline
$ \omega _{max }$ & 0.262567 & 0.260232 & 0.256539 & 0.251857 & 0.2468 & 0.239897 & 0.233864 & 0.227172 & 0.226427 \\
\hline
$ P_{max }$ & $\text{1.29101}\times 10^{-6}$ & \text{1.17109}$\times 10^{-6}$ & \text{1.00037}$\times 10^{-6}$ &
\text{8.10007}$\times 10^{-7}$ & \text{6.26519}$\times 10^{-7}$ & \text{4.66743}$\times 10^{-7}$ &
\text{3.37486}$\times 10^{-7}$ & \text{2.38537}$\times 10^{-7}$ & \text{1.64995}$\times 10^{-7}$ \\
\hline
$ P_{\text{tot}}$ & \text{2.64246}$\times 10^{-7}$ & \text{2.36790}$\times 10^{-7}$ & \text{1.98349}$\times 10^{-7}$ & \text{1.56492}$\times 10^{-7}$ & \text{1.1734}$\times 10^{-7}$ & \text{8.44004}$\times 10^{-8}$ & \text{5.87576}$\times 10^{-8}$ & \text{3.99097}$\times 10^{-8}$ & \text{2.66294}$\times 10^{-8}$ \\
\hline
$ \eta$ & 41523.25 & 45517.54 & 52807.77 & 64511.41 & 82616.50 & 108524.97 & 148144.16 & 205803.54 & 306417.49 \\
\hline
\end{tabular}
\label{sparsescalar}
\end{table}
\begin{table}[!htp]
\centering
\caption{Values of $\omega_{max}$, $P_{max}$, $P_{tot}$, and $\eta$ for electromagnetic perturbation for various values of $\alpha$ for $\ell=1$ mode.}
\setlength{\tabcolsep}{-.2mm}
\begin{tabular}{|c|c|c|c|c|c|c|c|c|c|}
\hline
$ \alpha$ & 0.2 & 0.4 & 0.6 & 0.8 & 1.0 & 1.2 & 1.4 & 1.6 & 1.8 \\
\hline
$ \omega _{\max }$ & 0.287926 & 0.285118 & 0.280054 & 0.274884 & 0.268825 & 0.262115 & 0.255269 & 0.241694 & 0.234996 \\
\hline
$ P_{\max }$ & 5.36194$\times 10^{-7}$ & 4.83117$\times 10^{-7}$ & 4.08259$\times 10^{-7}$ &
$ 3.25856\times 10^{-7}$ & 2.47758$\times 10^{-7}$ & $1.81044\times 10^{-7}$ &
1.28218$\times 10^{-7}$ & 8.84317$\times 10^{-8}$ & 6.01674$\times 10^{-8}$ \\
\hline
$P_{tot}$ & 1.13948$\times 10^{-7}$ & 1.01417$\times 10^{-7}$ & 8.40331$\times 10^{-8}$ & 6.53514$\times 10^{-8}$ & 4.81631$\times 10^{-8}$ & 3.39774$\times 10^{-8}$ & 2.31655$\times 10^{-8}$ & 1.53955$\times 10^{-8}$ & 1.00469$\times 10^{-8}$ \\
\hline
$ \eta$ & 115791.41 & 127573.91 & 148544.85 & 184019.17 & 238805.64 & 321821.27 & 447686.27 & 603892.31 & 874806.76 \\
\hline
\end{tabular}
\label{sparseem}
\end{table}
\section{Area spectrum for the LQG motivated black hole from adiabatic invariance}
In this section, we obtain the area spectrum of the LQG motivated $4D$ polymerized black hole from adiabatic invariance with the help of works \cite{br2011, qq2012}. First, we euclideanize the metric (\ref{metric2}) by using the transformation $t\rightarrow -i\tau$. It produces
\begin{eqnarray}\label{euclidean}
ds^{2}= \left(\frac{r-2M}{\sqrt{r^2+\alpha^2}}\right)d\tau^{2}+\frac{1}{\left(\frac{r-2M}{\sqrt{r^2+\alpha^2}}\right)}dr^{2}
+(r^2+\alpha^2)(d\theta^{2}+\sin^{2}\theta d\phi^{2}).
\end{eqnarray}
The radial null geodesics for the above metric are given by
\begin{equation}
\dot{r}=\pm i \frac{r-2M}{\sqrt{r^2+\alpha^2}}.
\end{equation}
Using the adiabatic invariant quantity given in \cite{br2011} and following the same procedure thereof, we have
\begin{equation}
I=\int p_{j}dq_{j}=\int_{0}^{M}\frac{dM}{T_H}.
\end{equation}
Here, $p_{j}$ is the momentum conjugate to the coordinate $q_{j}$ where j has two values, 0 and 1. We have $q_{0}=\tau$ and $q_{1}=r$. $T_{H}$ is the Hawking temperature given by Eq. (\ref{hawking}). Using above equation and Eq. (\ref{hawking}), we obtain
\begin{eqnarray}\nonumber
I&=&4 \pi \left(\frac{1}{2} M \sqrt{\alpha ^2+4 M^2}+\frac{1}{4} \alpha ^2 \log \left(\sqrt{\alpha ^2+4 M^2}+2 M\right)\right)\\\nonumber
&\approx&4 \pi \left(M^2+\frac{1}{8} \alpha ^2 (2 \log (4 M)+1)\right)\\
&=&4 \pi \left(\frac{1}{8} \alpha ^2 \left(\log \left(\frac{A}{\pi }\right)+1\right)+\frac{A}{16 \pi }\right).
\end{eqnarray}
Using the quantization rule of Bohr-Sommerfeld $I= 2\pi n$ with $n = 0,1,2,...$ we obtain
\begin{equation}
A_n=2 \pi \alpha ^2 W\left(\frac{e^{\frac{4 n}{\alpha ^2}-1}}{2 \alpha ^2}\right),
\end{equation}
where $W(z)$ is the Lambert W function. The area spectrum is given by
\begin{equation}
\Delta A=A_{n}-A_{n-1}.
\end{equation}
We observe that the area spectrum of the black hole is significantly different from that of the Schwarzschild black hole. The above equation reduces to the Schwarzschild case in the limit $\alpha \rightarrow 0$. Above equation along with Eq. (\ref{pl}) shows that the power spectrum for the LQG-motivated black hole is quantized and the quantization rule is different from the Schwarzschild case.
\section{conclusions}
In this article, we have used LQG motivated $4D$ polymerized black hole to study shadow, quasinormal modes, greybody bounds, and Hawking sparsity in the background spacetime. Due to quantum gravity correction, the black hole becomes regular. This was confirmed by finite values of scalar invariants everywhere. To calculate the shadow radius of the black hole, we first write down the corresponding Lagrangian for the metric and find out the differential equation of motion. There, we get the potential that dictates the motion of a particle. Imposing conditions on the potential, first, and second derivatives of the potential, we get the analytical expressions of the radius of the photon sphere $r_p$ and then, the radius of the black hole shadow $R_s$. The analytical expressions clearly show that the quantum correction impacts photon and shadow radii. To have qualitative as well as quantitative idea of the impact of the quantum correction on them, we plot $r_p$ and $R_s$ against $\alpha$ in Fig. (\ref{radius}) and we give numerical values of $r_p$ and $R_s$ for different values of the LQG parameter in Table (\ref{shadowradius}). Fig. (\ref{radius}) and Table (\ref{shadowradius}) show that both radii increase as we increase the value of $\alpha$ and their nature of variation against $\alpha$ is similar. We, then, plot shadows of the quantum corrected black hole for various values of $\alpha$ in Fig. (\ref{shadow}). Fig. (\ref{radius}), Fig. (\ref{shadow}), and Table (\ref{shadowradius}) conclusively show that quantum correction has a significant impact on the shadow of the black hole.\\
Next, we study the quasinormal modes of the black hole for two types of perturbations: scalar and electromagnetic using the $6th$ order WKB method. We plot the effective potential in Fig. (\ref{vfig}) with respect to normal coordinate r and briefly discuss the qualitative nature of the potential. Then, quantitative values of quasinormal modes for scalar and electromagnetic perturbations are given in Table (\ref{QNMS}) and Table (\ref{QNMEM}). In Fig. (\ref{qnmrefig}), we have plotted the real part of quasinormal frequency against $\alpha$ for various values of angular momentum $\ell$. However, the variation of the oscillation frequency with respect to $\alpha$ is small. In Fig. (\ref{qnmimfig}), we have shown the variation of the imaginary part of quasinormal modes with respect to $\alpha$ for different $\ell$. We can infer from them that the oscillation frequency of GWs decreases as we increase the value of $\alpha$, but increases with an increase in angular momentum. We also observe that the damping rate decreases with $\alpha$, but increases with $\ell$. In Fig. (\ref{qnmimrefig}), we compare real and imaginary parts of quasinormal modes for both oscillations. We observe that the oscillation frequency as well as damping rate is larger for scalar perturbation than electromagnetic perturbation. In Fig. (\ref{qnmorderfig}), we show the convergence of the WKB method for various $(n,\ell)$ pairs. It shows that when $n<\ell$, quasinormal frequency fluctuates even for higher order. In the next section, we show the ringdown waveform for both perturbations using the time domain integration method in Figs. (\ref{ringingalpha}, \ref{ringingl}). We observe from Fig. (\ref{ringingalpha}) that the frequency, as well as decay rate, decreases as we increase the parameter value of $\alpha$. We conclude from Fig. (\ref{ringingl}) that the frequency as well as the decay rate increases as we increase $\ell$. These conclusions and the conclusions drawn from Tables (\ref{QNMS}, \ref{QNMEM}) are consistent.\\
Then, we calculate Hawking temperature and greybody bounds for the quantum-corrected black hole. We observe that the Hawking temperature decreases with an increase in $\alpha$. We calculate analytical expressions of greybody bounds. It shows that greybody bounds for electromagnetic perturbations do not depend on $\alpha$. Fig. (\ref{gbfig1}) shows the dependence of greybody bounds for scalar perturbation on $\alpha$. We observe that though the probability of detecting Hawking radiation at spatial infinity increases with the LQG parameter, the variation of probability is very small with respect to $\alpha$. In Fig. (\ref{gbfig}), we have plotted greybody factors for both perturbations with different angular momentum. We infer that the transmission probability of Hawking radiation decreases with $\ell$.\\
Next, we study the power spectrum and sparsity of Hawking radiation. Our study finds that the maximum value of the power spectrum decreases and frequency where the power spectrum is maximum shifts towards the left as we increase the value of $\alpha$. It is also observed that the total power emitted decreases as we increase the value of the LQG parameter. We, then, study the sparsity of Hawking radiation using a dimensional less quantity $\eta$. Quantitative values of $\eta$ are given in Table (\ref{sparsescalar}) and Table (\ref{sparseem}). Our study shows that the radiation becomes more sparse i.e., the time gap between successive radiation quanta becomes larger as we increase the LQG parameter. At the same time, we observe that Hawking radiation for electromagnetic perturbation is more sparse than scalar perturbation. Finally, with the help of the Bohr-Sommerfeld quantization rule, we obtain the area spectrum for the quantum-corrected black hole. LQG parameter $\alpha$ significantly impacts the area spectrum. We hope that in the future, we will have sufficient experimental results which will help us decide the fate of quantum gravity.\\
\\
\textbf{Declaration of competing interests}\\
The author declares that the work is not influenced by any competing financial interest or personal relationship.

\end{document}